\documentclass[aip, apl, 10pt, twocolumn,unsortedaddress]{revtex4-1}
\usepackage{graphicx}

\begin{document}

\title{Tuning the Nonlinear Response of Coupled Split-Ring Resonators}

\author{Kirsty E. Hannam}
\email{kirsty.hannam@anu.edu.au}
\author{David A. Powell}
\author{Ilya V. Shadrivov}
\author{Yuri S. Kivshar}

\affiliation{Nonlinear Physics Center, Research School of Physics and Engineering,\\ The Australian National University, Canberra ACT 0200, Australia}

\begin{abstract}
We introduce the concept of controlling the nonlinear response of the metamaterial by altering its internal structure. We experimentally demonstrate tuning of the nonlinear response of two coupled split-ring resonators by changing their mutual position. This effect is achieved through modification of the structure of the coupled resonant modes, and their interaction with the incident field. By offsetting the resonators we control the maximum currents through the nonlinear driving elements, which affects the nonlinear response of the system.
\end{abstract}

\maketitle

The concept of engineering the electromagnetic properties of a material to possess a negative refractive index was introduced by Veselago in 1968~\cite{Veselago1968}. This was experimentally confirmed by Smith \textit{et al.} in 2000~\cite{Smithetal2000}, using a metamaterial combining split-ring resonators (SRRs) to provide negative permeability~\cite{Pendryetal1999}, and thin metal wires to provide negative permittivity. More recently, metamaterials composed of SRRs have been used to produce optical activity~\cite{Lietal2008}, offering an alternate route to negative refraction~\cite{Pendry2004}.

The effective properties of these metamaterials depend both on the properties of the individual SRRs, and on the arrangement of the SRRs in the lattice, as strong interactions occur between the rings in a lattice due to their complex near-field patterns. This allows us to significantly control the properties of the metamaterial by changing the internal structure of the composite material~\cite{Lapineetal2009}. In order to understand the resulting responses of the metamaterial, we need to understand the interactions between the individual SRRs. This is a major motivation behind studies involving a change of the relative positions of two SRRs, including rotating them~\cite{Powelletal2011}, and shifting them with respect to each other~\cite{Powelletal2010}.

The linear properties of individual metamaterial elements can also be controlled by an external signal, by inserting {\em nonlinear inclusions} such as diodes~\cite{Lapineetal2003} or semiconductors~\cite{Degironetal2007}  into the rings. Many tunable inclusions can also respond to the high-frequency incident wave, thus they form the basis of \emph{nonlinear metamaterials}, which typically have power-dependent resonant frequencies\cite{Shadrivovetal2006, Powelletal2007,Larouche2010,Huang2010}. The nonlinear response of metamaterials can be much stronger than that of natural materials, due to their resonant nature, and the local field enhancement which occurs at certain ``hot spots''.

The strength of the local field depends on the design of the resonator, and the choice of an optimal location to place the nonlinear element. However it also depends strongly on the coupling of the wave to the external field, which along with losses determines the quality factor of the resonator. Since modifications of the lattice parameters also influence the quality factor of the resonances~\cite{Powelletal2010}, they should affect the local-field enhancement. In this Letter we experimentally demonstrate \emph{control of the nonlinear response} of two broadside-coupled SRRs by modifying the offset between them, and explain this response by studying how shifting the resonances changes the maximum currents in the resonators.

\begin{figure}[tb]
	\centering
		\includegraphics[width=0.9\columnwidth]{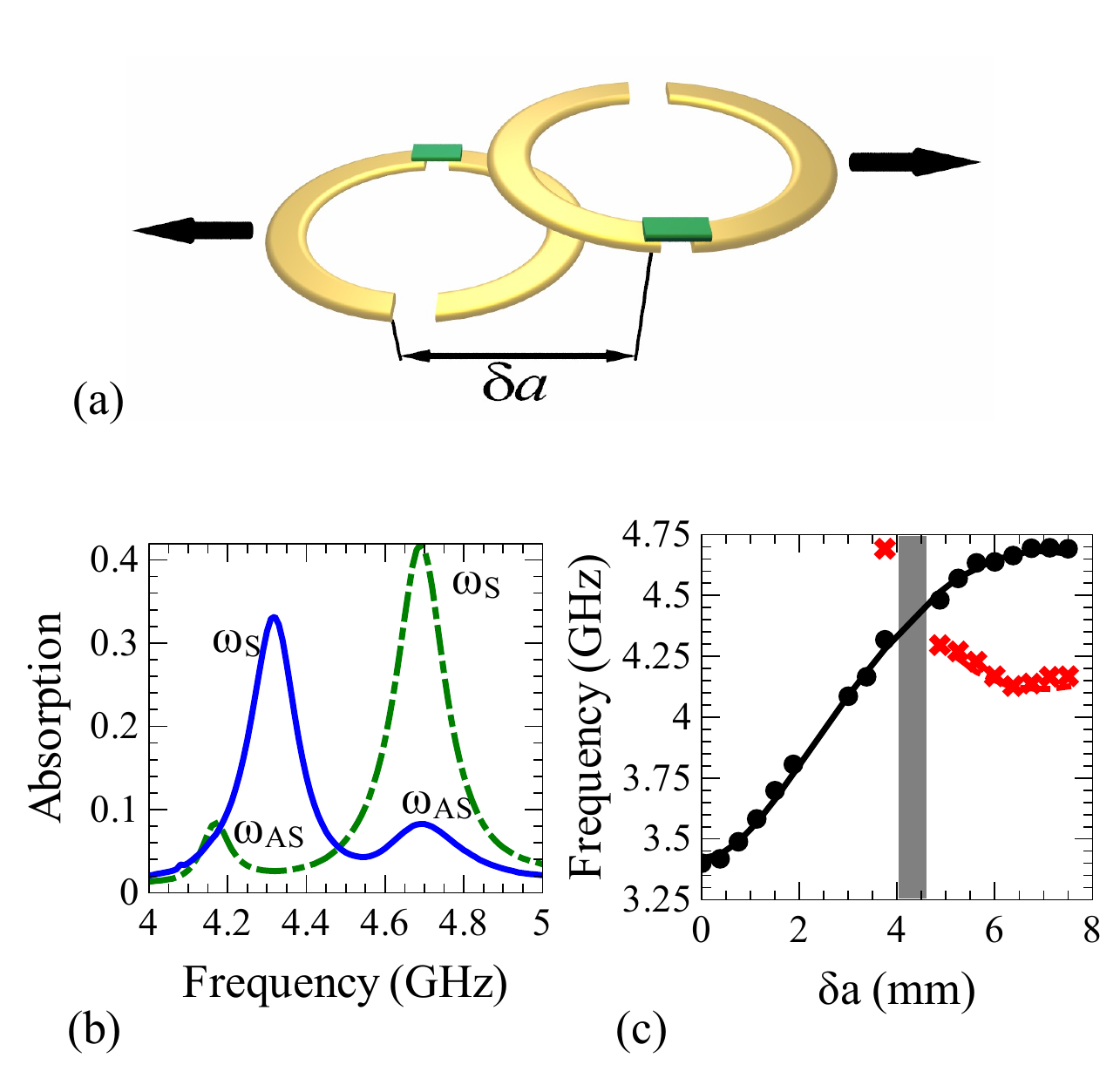}
	\caption{(a) Schematic of the lateral shifting of SRRs with inserted diodes; $\delta a$ is a shift between two rings. (b) The measured absorption curves for $\delta a= 3.75$\,mm (blue solid) and 7.5\,mm (green dashed), at -20\,dBm input power. Symmetric ($\omega_{S}$) and antisymmetric ($\omega_{AS}$) modes are highlighted. (c) $\omega_{S}$ (black circle/solid) and $\omega_{AS}$ (red cross/dashed), as determined both experimentally (markers) and numerically (lines) for the linear case.}
	\label{fig:Resonances}
\end{figure}

To start with we perform microwave experiments with a pair of SRRs, having an offset $\delta a$ between their centers, as shown schematically in Fig.~\ref{fig:Resonances}(a). The copper rings are 3.75\,mm in outer radius, 3.25\,mm in inner radius, and have a gap width of 1\,mm. These rings are printed on opposite sides of 1.6\,mm thick FR4 circuit board.  Each ring has a second gap of 0.4\,mm opposite the initial gap, across which a Skyworks SMV1405-079 series diode is soldered. The samples are placed inside a WR-229 rectangular waveguide, oriented such that the incoming magnetic field is perpendicular to the loops.

The transmission and reflection from the sample are measured using a Rohde and Schwarz ZVB-20 vector network analyzer. We measure the absorption spectrum, which describes the frequency of the maximum current distribution in the rings, given by $1 - |S_{21}|^2 - |S_{11}|^2$, where $S_{21}$ and $S_{11}$ are the transmission and reflection coefficients respectively. The absorption curves are measured experimentally for the values of $\delta a$ between 0 and 7.5\,mm, in 0.375\,mm increments, at powers -20\,dBm and 15\,dBm. Numerical results are calculated using CST Microwave Studio. Each diode is represented by a series RLC circuit with 2.67\,pF capacitor, 0.8$\Omega$ resistor and 0.7\,nH inductor. This circuit describes the diode in the linear regime~\cite{Datasheet}.

Figure~\ref{fig:Resonances}(b) shows the measured absorption curves for $\delta a$ equal to 3.75\,mm, and 7.5\,mm, at -20\,dBm. There are seen clearly two resonant modes, symmetric ($\omega_S$) and antisymmetric ($\omega_{AS}$), which are defined by the relative directions of the currents on the rings. These modes are both tunable through a change of $\delta a$~\cite{Lapineetal2009, Powelletal2010}, which is evident here, as when $\delta a = 3.75$\,mm $\omega_S$ ($\omega_{AS}$) is the lower (higher) frequency, while when $\delta a = 7.5$\,mm $\omega_S$ ($\omega_{AS}$) has increased (decreased) to become the higher (lower) frequency. Both $\omega_S$ and $\omega_{AS}$  are found and plotted as a function of $\delta a$ in Fig.~\ref{fig:Resonances}(c) (markers), along with the numerical equivalents (lines). When they overlap, it is difficult to accurately distinguish them in the data. Therefore the experimental values for $\delta a$ in the shaded regions have been excluded in all figures. We see excellent agreement between experiment and numerics. Consistent with the findings in Ref.~\onlinecite{Powelletal2010}, there are strong responses in both the modes, with $\omega_S$ increasing with an increase in $\delta a$.

\begin{figure}[tb]
	\centering
		\includegraphics[width=0.9\columnwidth]{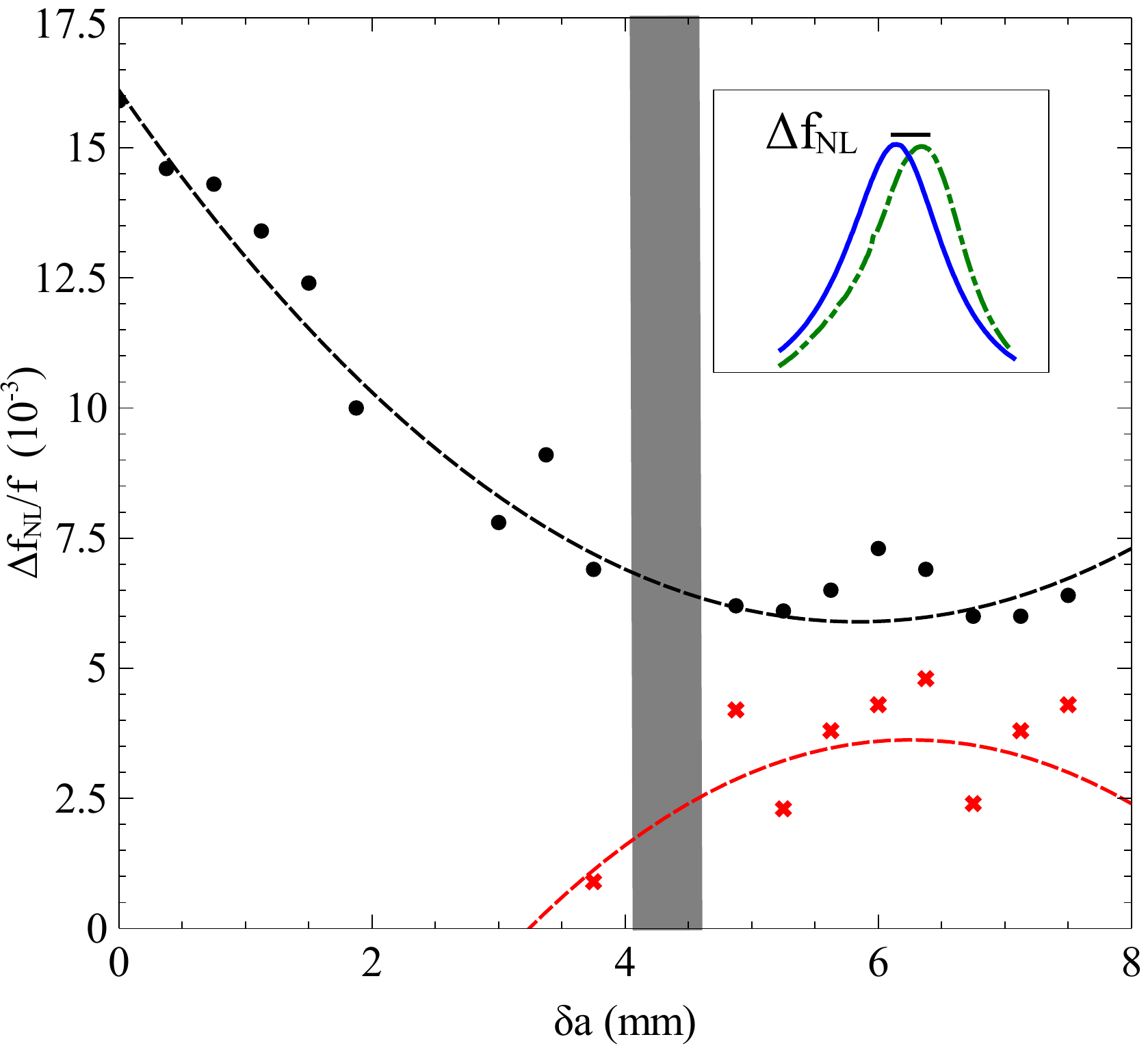}
	\caption{Relative nonlinear shift in $\omega_S$ (black circles) and $\omega_{AS}$ (red crosses) for shifts ($\delta a$). The dashed lines are second order polynomial fits to the data. The inset shows the nonlinear shift of the absorption at the symmetric mode when $\delta a = 3.75$\,mm. The solid blue line is at -20\,dBm, and the green dashed line at 15\,dBm.}
	\label{fig:Nonlinear}
\end{figure}

As seen in Fig.~\ref{fig:Resonances}(b), the symmetric mode is much more strongly excited than the antisymmetric mode. In fact, the antisymmetric mode is not visible for low values of $\delta a$, as seen in Fig.~\ref{fig:Resonances}(c), due to weak coupling to the incident wave. However, we have enough points to see that this mode is still tunable via the lateral shifting. It is also important to note that the first two measurements we have for this mode are only experimental, as it was still unobservable in the numerics.

In a single SRR, the resonant frequency is determined by $\omega = 1/\sqrt{LC}$, where $L$ is inductance, and $C$ is capacitance. Introducing the diode adds a nonlinear capacitance $C_{V}$ in series with the capacitance of the SRR. By adding the second SRR, we introduce coupling between the rings, which causes a split in the resonant frequency, resulting in the two previously discussed modes, $\omega_{S}$ and $\omega_{AS}$. By changing the intensity of the incoming microwaves, we are causing a nonlinear change in the capacitance of the diodes, which shifts the resonant frequencies by changing the effective capacitance of the system. This results in us having an added degree of freedom for manipulating our system.

In our structure the nonlinear response is quantified by a shift of the resonant frequency with increasing input power from -20\,dBm to 15\,dBm. The relative difference between the low- and high-power resonant frequencies for each value of $\delta a$ is shown in Fig.~\ref{fig:Nonlinear}. The inset in Fig.~\ref{fig:Nonlinear} shows the absorption of the symmetric peak for both powers when $\delta a = 3.75$\,mm. We observe a significant change in the nonlinear frequency shift for both modes as a function of $\delta a$. It is clear that the nonlinear shift for $\omega_{S}$ is much stronger when the rings are closer together. We do not have enough data points to meaningfully describe a trend for $\omega_{AS}$.

\begin{figure}[b]
	\centering
		\includegraphics[width=0.9\columnwidth]{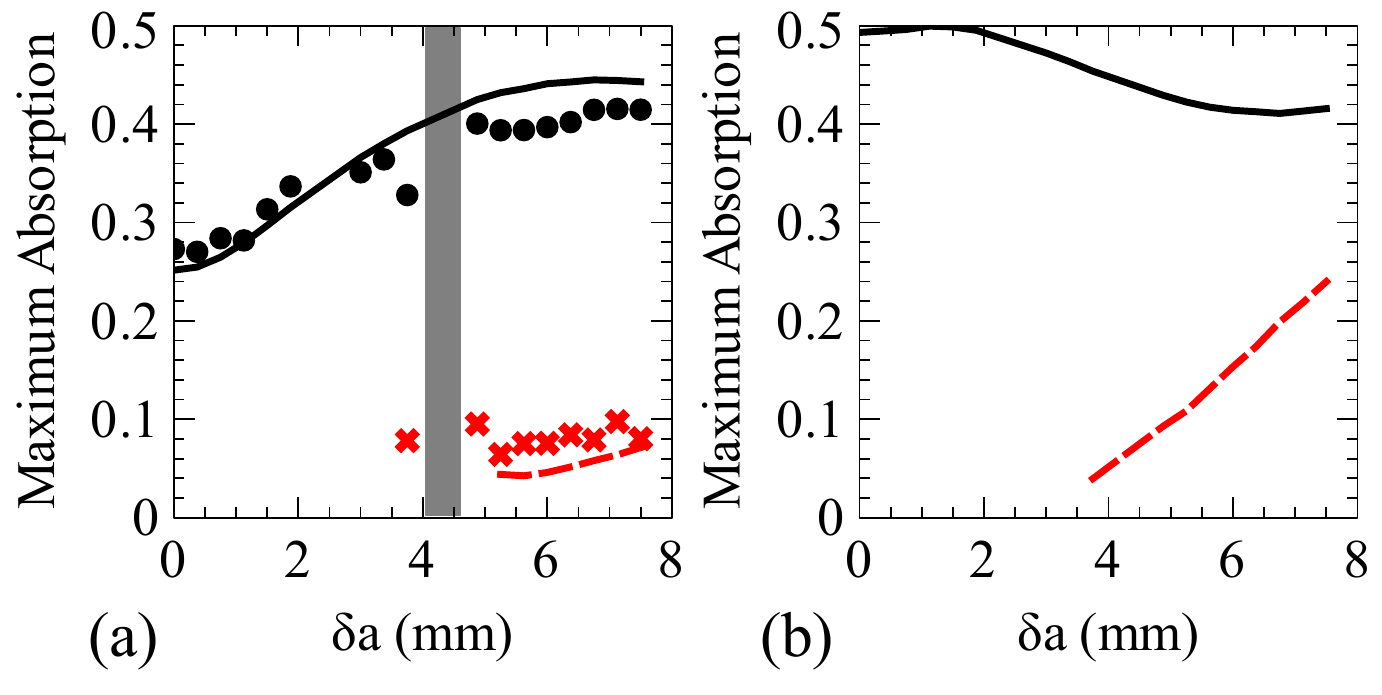}
	\caption{(a) Experimental (markers) and numerical (lines) maximum absorptions for both symmetric (black circle/solid) and antisymmetric (red cross/dashed) modes. (b) Numerically calculated absorption coefficients at the resonant frequencies for the case of a lossless substrate.}
	\label{fig:Absorptions}
\end{figure}

We can explain the nonlinear tuning by looking both experimentally and numerically at some of the linear aspects of the system. As the nonlinear shift is due to the currents excited in the rings (which also leads to absorption), it would be expected that the higher the absorption, the larger the nonlinear response. Therefore the maximum absorption for each value of $\delta a$ is plotted in Fig.~\ref{fig:Absorptions}(a) for both modes, showing good agreement between experimental and numerical results. However, by comparing this figure with Fig.~\ref{fig:Nonlinear}, for the symmetric mode we see \emph{increasing absorption} with \emph{decreasing nonlinear shift}. This effect appears because by offsetting the rings we not only modify the current amplitudes in the rings, but also change the field distribution in the substrate. As a result, the total absorption presented in Fig.~\ref{fig:Absorptions}(a) contains contribution from losses in both metal and dielectric. At the same time, the nonlinear response is caused by currents in the metal which flow through the diode. The losses within the FR4 substrate increase with $\delta a$ due to increased field confinement within the board, and change in these losses dominate changes of the absorption in the metal. In Fig.~\ref{fig:Absorptions}(b) we show the results of a simulation which neglects dielectric losses. We can see that we now have agreement with the trend in the nonlinear response. The difference in trends between Figs.~\ref{fig:Absorptions}(a) and (b) confirms the influence of the dielectric losses.

\begin{figure}[tb]
	\centering
		\includegraphics[width=\columnwidth]{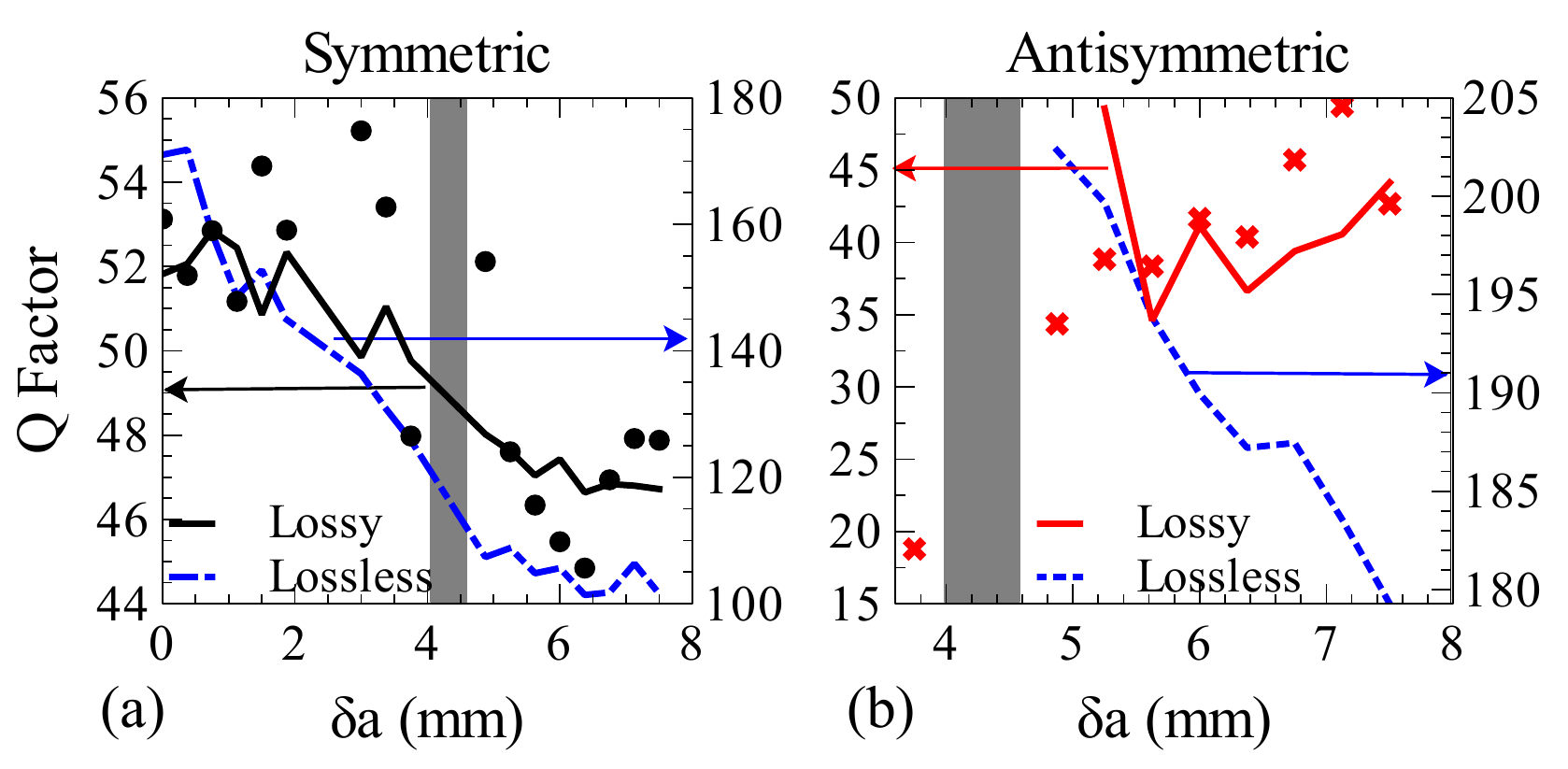}
	\caption{Experimental (markers) and numerical (lines) $Q$ factors for the (a) symmetric mode, and (b) antisymmetric mode. The red (solid) lines correspond to the lossy case, while the blue (broken) lines correspond to the lossless case. The arrows indicate the relevant axis.}
	\label{fig:Q}
\end{figure}

The amplitude of the currents in the system depends on the $Q$ factor of the resonator, so we would expect to see some correlation between the changing nonlinear shift, and the Q factor. In Fig.~\ref{fig:Q}, we have plotted the Q factor as a function of $\delta{a}$ for both modes. The solid lines for the symmetric and antisymmetric mode is the Q factor in the numerical lossy case. By comparing this with Fig.~\ref{fig:Nonlinear} we see that we have a direct correlation due to the change in dielectric losses. The additional finely dashed curves show the case when dielectric losses are neglected. This indicates that for both modes radiation losses increase with increasing offset, due to increased coupling to the waveguide modes. For the symmetric mode, radiation losses are dominant, thus increasing them decreases the current amplitude and hence the nonlinear shift. However for the anti-symmetric mode, the coupling is initially very low, and for low offsets very little energy can accumulate in the resonant mode. Therefore increasing the coupling increases the current amplitude and the nonlinear shift.

\begin{figure}[tb]
	\centering
		\includegraphics[width=0.9\columnwidth]{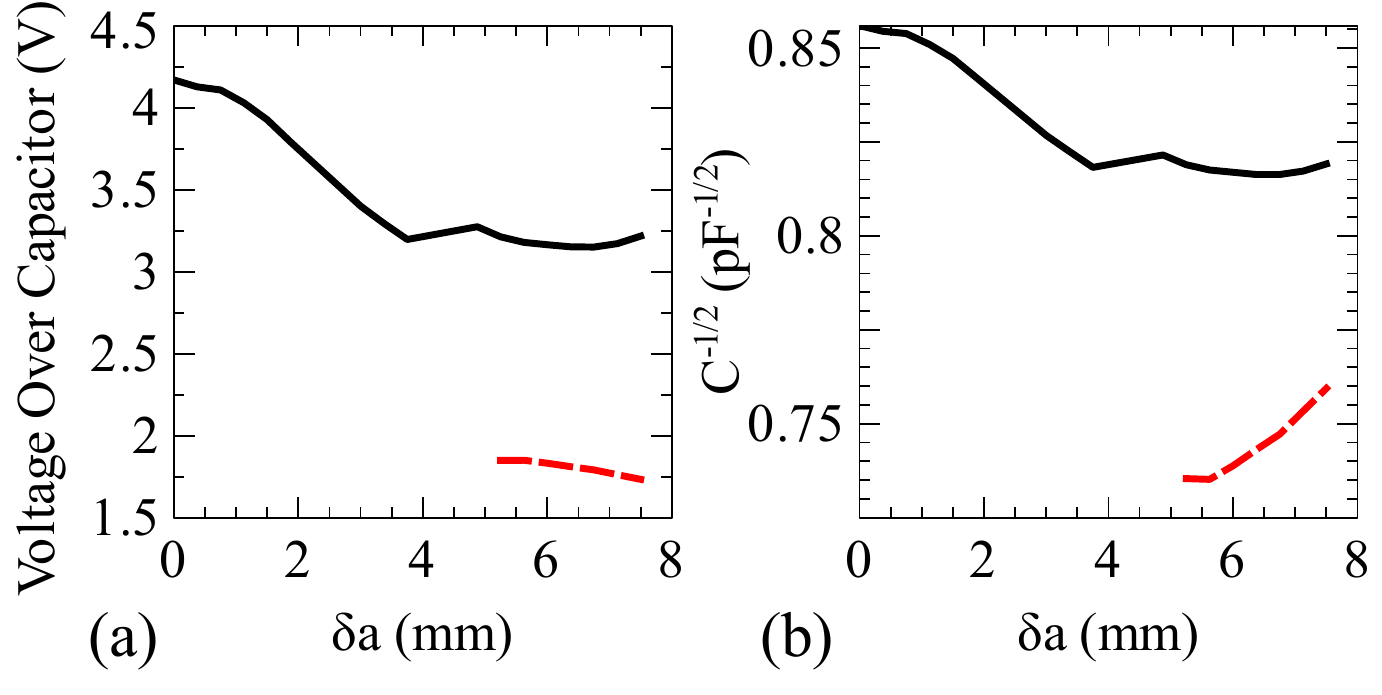}
	\caption{(a) Calculated voltage over the capacitive component of the diode. (b) Resulting capacitance (presented as $1/\sqrt{C}$).}
	\label{fig:Voltage}
\end{figure}

The physics of the nonlinear shift of the resonant frequency was studied in Ref.~\onlinecite{Powelletal2007}, where it was shown that the rectified AC voltage across the varactor provides a self-biasing mechanism. By increasing the amplitude of the electromagnetic wave, we increase alternating voltages in the SRR, generating a larger DC bias voltage, reducing the varactor capacitance. To study this effect, we performed numerical simulations, and monitored the voltages on each RLC circuit representing a varactor. As a larger voltage leads to a stronger decrease in the varactor capacitance, we would expect to see a similar trend between the voltages recorded at the resonant frequency, and the nonlinear shift. We note that the voltage across the capacitance, and not the whole RLC circuit should be calculated. Corresponding DC voltage at the resonant frequency is plotted for both resonant modes in Fig.~\ref{fig:Voltage}(a).

By comparing Figs.~\ref{fig:Nonlinear},~\ref{fig:Absorptions}(b), and~\ref{fig:Voltage}(a), we can see an overall correlation, where as $\delta a$ increases there is a decreasing trend in the nonlinear response, the maximum lossless absorptions, and the calculated voltages, for the symmetric mode. This enables us to explain the nonlinear response of the system in a meaningful way.

As the nonlinear response should be directly proportional to the inverse of the square root of the capacitance, we then calculated this capacitance using the known voltage across the diode.  From the data sheets for the diode, we can obtain the capacitance ($C_{V}$) for a given applied voltage ($V_{R}$) using the equation~\cite{Datasheet}
\begin{equation}
C_{V} = \frac{C_{j0}}{(1 + V_{R}/ V_{j})^{M}} + C_{P},
\end{equation}
where $C_{j0}$ is the zero bias junction capacitance, $V_{j}$ is the junction potential, $M$ is the grading coefficient, and $C_{P}$ is the package capacitance. For the specific diode series we have used, these values are $C_{j0} = 2.92$\,pF, $V_{j} = 0.68$\,V, $M = 0.41$, and $C_{P} = 0.05$\,pF. The inverse of the square root of the resulting capacitance is shown in Fig.~\ref{fig:Voltage}(b), as a function of $\delta a$. As expected, this plot closely resembles the plots in Fig.~\ref{fig:Voltage}(a) and Fig.~\ref{fig:Nonlinear}. The dependence of the diode capacitance on $\delta a$ also changes the response of the system to linear tuning of the resonant frequencies, as the resonant frequency is dependent on the combined capacitance of the diode and SRR.

In conclusion, we have demonstrated that by shifting two coupled SRRs relative to each other, we can significantly control the nonlinear properties of both the symmetric and antisymmetric resonant modes in the system. We have also found that the tuning of the nonlinear response can be explained and predicted by studying the voltages generated across the diodes. We expect that our results will stimulate further work in controlling and designing nonlinear properties of metamaterials.

We acknowledge a financial support from the Australian Research Council.

\end{document}